\documentclass[showpacs,preprintnumbers,amsmath,amssymb,prb,twocolumn]{revtex4}

\usepackage{graphicx}               
\usepackage{psfrag}
\usepackage{dcolumn}                
\usepackage{bm}                     

\newcommand{\ba}{\begin{array}}
\newcommand{\ea}{\end{array}}
\newcommand{\bc}{\begin{center}}
\newcommand{\ec}{\end{center}}

\newcommand{\eps}{\epsilon}

\newcommand{\tr}{\textrm{ tr}}

\newcommand{\UA}{\uparrow}
\newcommand{\DA}{\downarrow}

\newcommand{\phd}{^{\phantom{\dagger}}}
\newcommand{\SC}{{\cal S}}
\newcommand{\Dint}{\int {\cal D}[\SC]\;}
\newcommand{\YY}{{\cal Z}}

\begin{document}

\newcommand{\Eq}[1]{{Eq.~(\ref{#1})}}
\newcommand{\EQ}[1]{{Equation~(\ref{#1})}}
\newcommand{\av}[1]{{\left<{#1}\right>}}
\newcommand{\E}{{\textrm{e}}}
\newcommand{\note}[1]{.\newline\marginpar{\LARGE\bf!}{\bf #1}\newline}

\newcommand{\cdag}{c^\dagger}
\newcommand{\cnod}{c^{\phantom{\dagger}}}
\newcommand{\adag}{a^\dagger}
\newcommand{\anod}{a^{\phantom{\dagger}}}
\newcommand{\ctdag}{\tilde c^\dagger}
\newcommand{\ctnod}{\tilde c^{\phantom{\dagger}}}

\title{Magnetic Polarons in the 1D FM Kondo Model} 

\author{Winfried Koller, Alexander Pr\"ull, Hans Gerd Evertz,
  and Wolfgang von der Linden}
 \affiliation{Institut f\"ur Theoretische Physik,
   Technische Universit\"at Graz, Petersgasse 16, A-8010 Graz,
   Austria.}
\email{koller@itp.tu-graz.ac.at}

\date{January 20, 2003}

\begin{abstract}
  The ferromagnetic Kondo model with classical corespins
  is studied via unbiased Monte-Carlo simulations.
  We show that with realistic parameters for the manganites
  and at low temperatures, the double-exchange mechanism does not lead
  to phase separation in one-dimensional chains but rather stabilizes
  individual ferromagnetic polarons.
  Within the ferromagnetic polaron picture, the pseudogap in the one-particle
  spectral function $A_k(\omega)$ can easily be explained.
  Ferromagnetic polarons also clear up a seeming failure of the double-exchange
  mechanism in explaining the comparable bandwidths in the ferromagnetic
  and paramagnetic phase.
  For our analysis,  
  we extend a simplified model, the finite temperature uniform hopping
  approach (UHA), to include polarons. 
  It can easily be evaluated numerically and provides a simple quantitative
  understanding of the physical features of the ferromagnetic Kondo model.  
\end{abstract}



\maketitle

\section{Introduction}                                  \label{sec:intro}

Manganese oxides such as
La$_{1-x}$Sr$_x$MnO$_3$  and La$_{1-x}$Ca$_x$MnO$_3$ have been attracting
considerable attention since the discovery of colossal magnetoresistance
(CMR)~\cite{proceedings98,Nagaev:book}.
These materials crystalize in the
perovskite-type lattice structure where the crystal field breaks the symmetry
of the atomic wave function of the manganese $d$-electrons. The energetically
lower  $t_{2g}$ levels are occupied by three localized electrons. Due
to a strong Hund coupling their spins are aligned, forming a localized
corespin with $S=3/2$. The electron configuration of the Mn$^{3+}$ ions is
$t_{2g}^3e_g^1$, whereas for Mn$^{4+}$ ions the $e_g$ electron is missing. Due
to a hybridization of the $e_g$ wave function with the oxygen $2p$ orbitals,
the $e_g$ electrons are itinerant and can move from an Mn$^{3+}$ ion to a
neighboring Mn$^{4+}$ via a bridging O$^{2-}$.
The interplay of various physical ingredients
such as the strong Hund coupling ($J_H$) of the itinerant electrons to
localized corespins, Coulomb correlations, and electron-phonon coupling leads
to a rich phase diagram including antiferromagnetic insulating, ferromagnetic
metallic and charge ordered domains.
The dynamics of the charge carriers moving in the spin and orbital background
shows remarkable dynamical features~\cite{horsch99,bala02}.

Since full many-body calculations for a realistic model, including all
degrees of freedom, are not possible yet, several approximate studies of
simplified models have been performed in order to unravel individual pieces
of the rich phase diagram of the manganites. 
The electronic degrees of freedom are generally treated by a Kondo lattice
model which in the strong Hund coupling limit is commonly referred to as
the double-exchange (DE) model, a term introduced by Zener~\cite{zener51}.
In addition, the correlation of the itinerant $e_g$ electrons is well
described by a nearest neighbor (n.n.)\ Coulomb interaction.
The on-site Hubbard term merely renormalizes the already strong Hund
coupling.
For the Kondo model with quantum spins it is still impossible to derive
rigorous numerical or analytical results.
If the S=3/2 corespins are treated classically, however, the model can be
treated by unbiased Monte Carlo techniques.
The impact of quantum spins on the electronic properties has been studied
in Ref.~\onlinecite{EdwardsI,Nolting01,Nolting03}.
It appears that quantum effects are important for (S=1/2) corespins or at
$T=0$.
For finite temperature and S=3/2, classical spins present a reasonable
approximation.

Elaborate Monte Carlo (MC) simulations for the FM Kondo model with
classical $t_{2g}$ corespins have been performed by Dagotto
{\em et al.}~\cite{dagotto98:_ferrom_kondo_model_mangan,
yunoki98:_static_dynam_proper_ferrom_kondo,yunoki98:_phase},
Yi {\it et al.}~\cite{Yi_Hur_Yu:spinDE}, and by
Furukawa {\it et al.}~\cite{furukawa98,Motome_Furukawa_3dDE}.
Static and dynamical properties of the model have been determined.
These studies revealed features (discontinuity of the mean electron density
as a function of the chemical potential, infinite compressibility)
which have been interpreted as signatures of phase-separation (PS).
PS has also been reported~\cite{Millis_PS} from computations based on a
dynamical mean field treatment of the DE model at $T=0$.
A phase diagram and critical exponents of the DE model have been determined
with a Hybrid MC algorithm\cite{Mayor_hybrid2001,Mayor_DE2001}.

In the manganites, the Hund coupling $J_H$ is much stronger than the
kinetic energy.
Consequently, configurations are very unlikely in which the
electronic spin is antiparallel to the local corespin.
The present authors have proposed an effective spinless fermion (ESF)
model~\cite{KollerPruell2002a} that takes effects of
antiparallel spin configurations into account via virtual excitations.
It has been demonstrated that the results of the ESF model are in excellent
agreement with those of the original Kondo model even for moderate values
of $J_H$.
This applies also to features which have been previously interpreted
as signatures of PS~\cite{yunoki98:_phase}.
Taking Coulomb interactions into account, PS has been
argued to lead to either small\cite{Sboychakov_02} or
large\cite{moreo_science_99} (nano-scale) clusters, which have been the
basis for a possible though controversial\cite{EdwardsI} explanation of
CMR\cite{dagotto01:review, Millis_III}. Moreover, lattice
distortions~\cite{allen99:_anti_jahn_teller_lamno,yunoki98:_phase_separ_induc_orbit_degrees}
are believed to play a crucial role for the CMR
effect~\cite{EdwardsI,millis96:polaronsI} and should also be included
in the model.

In this paper, we present a numerical and analytical study of the
1D ferromagnetic Kondo model with classical corespins.
We find that the correct physical interpretation of the features which
have been interpreted as PS in the one-dimensional model,
is rather given by ferromagnetic polarons,
i.e.\ small FM-regions with {\em one single} trapped charge-carrier,
compatible with exact diagonalization results for small
clusters~\cite{horsch99}.
This applies even without n.n.\ Coulomb repulsion invoked in
Refs~\onlinecite{moreo_science_99,Sboychakov_02}.
Energetically, there is no significant difference between polarons,
bi-polarons or even charge accumulations in the PS sense.
It is rather the entropy, which even near zero temperature
clearly favors polarons.
The polaron picture allows also a straight forward and obvious
explanation of the pseudogap, which has been previously observed in
the spectral
density~\cite{DessauI, DessauII, dagotto01:review, KollerPruell2002a}. 

In a previous paper~\cite{KollerPruell2002a} we introduced the uniform
hopping approximation (UHA), which replaces the influence of the random
corespins on the $e_g$ electron dynamics by an effective uniform
hopping process.
Essential physical features of the original model could be described even
quantitatively by UHA, while the configuration space, and hence the numerical
effort, was reduced by several orders of magnitude.
Besides the numerical advantage, UHA also allows the derivation of
analytical results in some limiting cases at $T=0$.

In Ref.~\onlinecite{KollerPruell2002b} we have extended UHA to finite
temperatures by allowing for thermal fluctuations of the uniform
hopping parameter.
By taking into account the density of corespin states,
it is possible to calculate thermodynamic quantities of one and
three-dimensional systems in the UHA.
The reliability of this finite-temperature UHA has been scrutinized by a
detailed comparison of the results for various properties of the
ferromagnetic Kondo model with unbiased MC data in 1D.

Here, we will generalize UHA to regimes, where a single
hopping parameter is not sufficient to describe the physics of the FM Kondo
model.
Particularly near half filling, a typical corespin configuration shows small
ferromagnetic domains (polarons) immerged in an antiferromagnetic background.
Therefore, two different UHA-parameters are necessary to model the impact of
the fluctuating corespins on the $e_g$ electron dynamics.

This paper is organized as follows.
In Sec.~\ref{sec:model} the model Hamiltonian is presented and
particularities of the MC simulation for the present model are outlined.
The general discussion of ferromagnetic polarons near half filling is
presented in Sec.~\ref{sec:FMP}.
Section~\ref{sec:UHA_FMP} develops a generalization
of the uniform hopping approach in order to treat FM polarons.
Polaronic features  in the spectral density are analyzed.
The key results of the paper are summarized in Sec.~\ref{sec:conclusion}.

\section{Model Hamiltonian and unbiased Monte Carlo}        \label{sec:model}

In this paper, we will concentrate solely on properties of the
itinerant $e_g$ electrons interacting with the {\it local} $t_{2g}$ corespins.
We also neglect the degeneracy of the $e_g$ orbitals.
The degrees of freedom of the $e_g$ electrons are then described by a
single-orbital Kondo lattice model~\cite{KollerPruell2002b}.
As proposed by de Gennes~\cite{gennes60},
Dagotto {\it et al.}~\cite{dagotto98:_ferrom_kondo_model_mangan,
dagotto01:review} and Furukawa~\cite{furukawa98}, the $t_{2g}$
spins~$\mathbf S_i$ are treated classically, which is equivalent to the
limit $S\to \infty$.
The spin degrees of freedom~$(\SC)$ are thus replaced by unit vectors
$\mathbf S_i$, parameterized by polar and azimuthal angles  $\theta_i$ and
$\phi_i$, respectively.
The magnitude of both corespins and $e_g$-spins is absorbed into the exchange
couplings.

\subsection{Effective Spinless Fermions}

It is expedient to use the individual $t_{2g}$ spin direction $\mathbf S_i$ as
the local quantization axis for the spin of the itinerant $e_g$ electrons at
the respective sites.
This representation is particularly useful for the $J_H\to\infty$ limit, but
also for the projection technique, which takes into account virtual processes
for finite Hund coupling.
As described in Ref.~\onlinecite{KollerPruell2002a}, the energetically
unfavorable states with $e_g$ electrons antiparallel to the local $t_{2g}$
corespins can be integrated out.
This yields the 1D effective spinless fermion model (ESF)
\begin{equation}                                       \label{eq:H}
  \hat H = -\sum_{<i,j>} t^{\UA\UA}_{i,j}\,
    \cdag_{i}\,\cnod_{j} - \sum_{i,j}
    \frac{t^{\UA\DA}_{i,j}\,t^{\DA\UA}_{j,i}}{2J_H}\, \cdag_{i}\cnod_{i}
    + J'\sum_{<i,j>} \mathbf S_i \cdot \mathbf S_j \;.
\end{equation}
The spinless fermion operators~$\cnod_{j}$ correspond to spin-up electrons
(relative to the {\em local} corespin-orientation) only.
The spin index has, therefore, been omitted.
With respect to a {\em global} spin-quantization axis the ESF
model~(\ref{eq:H}) still contains contributions from both spin-up and
spin-down electrons.

The first term in \Eq{eq:H} corresponds to the kinetic energy in
tight-binding approximation.
The modified hopping integrals $t^{\sigma,\sigma'}_{i,j}$
depend upon the $t_{2g}$ corespin orientation
\begin{equation}                                      \label{eq:modihop}
  t^{\sigma,\sigma'}_{i,j} \;=\;
  t_{0}\; u_{i,j}^{\sigma,\sigma'}\;,
\end{equation}
where the relative orientation of the $t_{2g}$ corespins at site $i$ and
$j$ enters via
\begin{equation}
 \begin{aligned}
    u^{\sigma,\sigma}_{i,j}(\SC)  &= \cos(\vartheta_{ij}/2)\;\E^{i\psi_{ij}}\\
    u^{\sigma,-\sigma}_{i,j}(\SC) &= \sin(\vartheta_{ij}/2)\;\E^{i\chi_{ij}}
  \end{aligned}\;.
\end{equation}
These factors depend on the relative angle $\vartheta_{ij}$ of
corespins $\mathbf S_i$ and $\mathbf S_j$ and on some complex phases
$\psi_{ij}$ and $\chi_{ij}$.

The second term in \Eq{eq:H} accounts for virtual hopping processes to
antiparallel spin--corespin configurations and vanishes in the limit
$J_H \to \infty$.
The last term is a small antiferromagnetic exchange of the corespins.

It should be noted that the unitary transformation to the local spin
quantization axis is not unique.
This fact can be exploited to eliminate the phase factors~$\psi_{ij}$ in 1D.
Then the n.n.\ hopping integrals are simply given by the real numbers
$\cos(\vartheta_{ij}/2)$.

\subsection{Grand Canonical Treatment}

We define the grand canonical partition function as
\begin{equation}                                           \label{eq:Y}
\begin{aligned}
  \YY &= \Dint\,\tr_c\, \E^{-\beta (\hat H(\SC)-\mu \hat{N})}\\
  \Dint &=  \prod_{i=1}^L\;\Big(\int_{0}^{\pi} d\theta_i\sin \theta_i
  \int_{0}^{2\pi} d\phi_i\Big)\;,
\end{aligned}
\end{equation}
where $\tr_c$ indicates the trace over fermionic degrees of freedom at
inverse temperature $\beta$, $\hat{N}$ is the operator for the total number
of $e_g$ electrons and $\mu$ stands for the chemical potential.
Upon integrating out the fermionic degrees of freedom, we obtain
the statistical weight of a corespin configuration $\SC$ that can
be written as
\begin{equation}                                  \label{eq:MC_weight}
    w(\SC) = \frac{\tr_c\, \E^{-\beta (\hat H(\SC)-\mu \hat{N})}}{\YY}\;.
\end{equation}

\EQ{eq:Y} is the starting point of Monte Carlo simulations of the Kondo
model~\cite{dagotto98:_ferrom_kondo_model_mangan}
where the sum over the classical spins is performed via importance sampling.
The spin configurations $\SC$ enter the Markov chain according to the
weight factor $w(\SC)$ that is computed via exact diagonalization of the
corresponding one-particle Hamiltonian in \Eq{eq:H}.
In the 1D case we have performed MC simulations in which spins in domains of
random lengths were rotated.
We have performed MC runs with 2000 measurements.
The skip between subsequent measurements was chosen to be some hundreds of
lattice sweeps reducing autocorrelations to a negligible level.

As previously shown\cite{KollerPruell2002b}, the spin-integrated one-particle
Green's function can be written as
\begin{equation}                                          \label{eq:MC_GF2}
  \sum_\sigma
  \ll a\phd_{i \sigma}; a^\dagger_{j \sigma} \gg_\omega
  = \Dint w(\SC) u_{ji}^{\UA\UA}(\SC)
    \ll c\phd_{i}; c^\dagger_{j} \gg^\SC_\omega\;,
\end{equation}
where $\ll c\phd_{i}; c^\dagger_{j} \gg^\SC_\omega$ is the
Green's function in local spin quantization.
It can be expressed in terms of the
one-particle eigenvalues $\eps^{(\lambda)}$ and the corresponding
eigenvectors $|\psi^{(\lambda)}\rangle$ of the Hamiltonian $\hat H(\SC)$:
\[
\ll \cnod_{i}; \cdag_{j}\gg^\SC_\omega
= \sum_\lambda \; \frac{\psi^{(\lambda)}(i) \;\psi^{*(\lambda)}(j)}
{\omega - (\eps^{(\lambda)} -\mu) + i0^+ }
\]
It should be pointed out that the one-particle density of states (DOS)
is independent of the choice of the spin-quantization.

\subsection{Uniform Hopping Approach}

The integral over the corespin states in the partition function (\ref{eq:Y})
can be evaluated approximatively by resorting to a uniform hopping approach
(UHA) \cite{KollerPruell2002b}. The key idea is to replace the impact of the
locally fluctuating corespins on the hopping amplitudes by some global
average quantity~$u$.
Then the Hamiltonian merely depends on one parameter, namely~$u$, and the
partition function can be written as the one-dimensional integral
\[
  \YY = \int_0^1 du\,\Gamma(u)\,\tr_c\, \E^{-\beta (\hat H(u)-\mu\hat N)}\;.
\]
The density of corespin states $\Gamma(u)$ can be obtained numerically
for 3D systems and analytically for 1D systems.

Obviously, this simplification assumes a uniform medium.
In order to cope with magnetic polarons, the UHA has to be generalized
and $\Gamma(u)$ should be replaced by a two-parameter density
$\Gamma(u_f,u_a)$, where $u_f (u_a)$ denotes the average hopping
within the (anti)ferromagnetic domains.
The details and results of such a generalization are the contents
of Sec.~\ref{sec:UHA_FMP} of this paper.

\section{Ferromagnetic Polarons}                                  \label{sec:FMP}

Near half filling of a single $e_g$ band, a tendency towards
phase separation has been observed in various studies.
It has been claimed that the system separates into FM domains of high
carrier concentration and AFM domains of low carrier concentration.
In the following we show that a different picture rationalizes
the 1D Monte Carlo results in the range $n\approx 0.7 - 1.0$.

\begin{figure}
  \psfrag{n}{\large $n_h$}
  \includegraphics[width=0.46\textwidth]{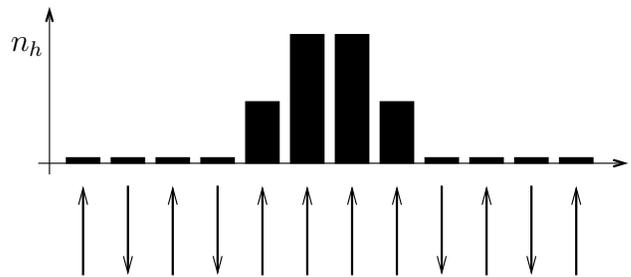}
  \caption{Idealized spin and hole-density configuration in a 1D Kondo
    chain at the critical chemical potential $\mu^*$.
    A FM domain of $L_f=4$ lattice sites is embedded in an AFM background.
    A single hole is localized in the FM domain giving rise to the depicted
    hole density (different from the schematic shape in Fig~4 of
    Ref.~\onlinecite{moreo99}).
  }
  \label{polaron.eps}
\end{figure}

We show that ferromagnetic polarons, i.e.\ {\em single} charge carriers
surrounded by small ferromagnetic spin-clouds, are formed when holes are
doped into the half filled $e_g$ band.
In order to model such a polaron in a one-dimensional system, we
take $L_f$ adjacent lattices sites to be in ferromagnetic order and
use $\Gamma_{\!L-L_f}(u)$ to account for the degeneracy of the remaining
spins.

First, we estimate the size $L_f$ of the FM polaron~\cite{allen01:sl_polaron}
using a simple polaron picture.
In this view the hole is confined in a perfectly FM
domain consisting of $L_f$ lattice sites and outside the domain the
system is in perfect AFM order (see Fig.~\ref{polaron.eps}).
The tight-binding energies in a potential well (FM domain) with infinite
barrier height are
\begin{equation}                                      \label{eq:tb_energies}
    \eps_\nu = -2\,
    \cos \big(\frac{\nu\;\pi}{L_f\!+\!1}\big)\;,\qquad\nu=1,\ldots,L_f\;.
\end{equation}
The energy difference between
a) a one-polaron state with perfect FM spins within the polaron and perfect
AFM order outside
and
b) perfectly antiferromagnetically ordered $t_{2g}$ spins
is given by
\[
  \Delta\eps_p = -2 \cos \Big(\frac{\pi}{L_f\!+\!1}\Big)
  + 2 J_{\text{eff}}\,(L_f-1)\;,
\]
where the first term accounts for the kinetic (delocalization) energy
of the hole in the potential well and the second term describes
the energy deficiency due to $(L_f-1)$ ferromagnetic bonds.
We have introduced the effective antiferromagnetic coupling $J_{\text{eff}}$,
given by
\[
  J_{\text{eff}} = J' + 1/(2 J_{H})
\]
near $n\approx 1$ (see Ref.~\onlinecite{KollerPruell2002a}).
For typical values $J_H=6$ and $J'=0.02$ we have
$J_{\text{eff}}\approx 0.1$.
Upon minimizing $\Delta\eps_p$ with respect to $L_f$, we obtain the optimal
size of the polaron, which in the present case lies between $L_f=3$ and
$L_f=4$.
If the FM domain contains $N>1$ charge carriers, the energy difference is simply
\[
  \Delta\eps_p(N) = -2 \sum_{\nu=1}^N\;\cos \Big(\frac{\nu\;\pi}{L_f\!+\!1}\Big)
  + 2 J_{\text{eff}}\,(L_f-1)\;.
\]
For $N=2$, the optimum bi-polaron size is $L_f\simeq 7$ and it increases to
$L_f\simeq 10$ for $N=3$ charge carriers.

\begin{figure}[h]
  \includegraphics[width=0.46\textwidth]{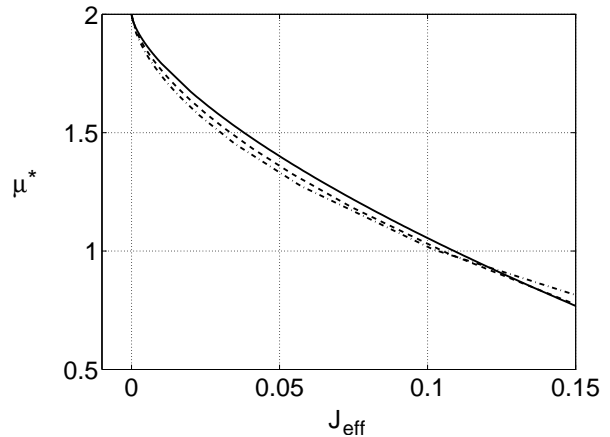}
  \caption{Critical chemical potential $|\mu^c|$ of the homogeneous model
    (solid line) from Ref.~\onlinecite{KollerPruell2002a} as compared to the
    chemical potential $\mu^*$ at which polarons (dash dotted) and bipolarons
    (dashed) start to enter the Kondo chain.}
  \label{muc_J.eps}
\end{figure}

Next we estimate the chemical potential $\mu^*$, at which  holes start
to populate the polaron states.
Apart from the energy of the antiferromagnetic $t_{2g}$ spins,
the total energy (at $T=0$) of the filled $e_g$ band is given in the grand
canonical ensemble by $-\mu^*\,L$.
By equating this energy to the total energy of the polaron we have
\[
  -\mu^*\,L = -\mu^*\,(L-1) + \Delta\eps_p(1)
\]
which yields the desired chemical potential
\[
  \mu^* = -\Delta \eps_p(1) = 2 \cos \big(\frac{\pi}{L_f\!+\!1}\big)
  - 2 J_{\text{eff}}\,(L_f-1)\;.
\]
The critical chemical potential is depicted in Fig.~\ref{muc_J.eps}.
Similar considerations yield that $\mu^*$ also approximately presents the
limiting chemical potential between the filled antiferromagnetic band and a
state with several single FM polarons, provided the polaron density is low,
i.e.\ as long as we have an antiferromagnetic background.
Consequently, at the chemical potential $\mu^*$ the electron density is
not fixed. The energy gain $\Delta \eps_p$ exactly balances the loss of
the chemical potenital $-\mu^*$.
This implies that the electron density has a discontinuity at $\mu^*$,
{\it i.e.} the compressibility of the electrons diverges
which has been previously interpreted as a consequence of PS tendencies
(see Fig. 3a in Ref.~\onlinecite{moreo_science_99} and Ref.~\onlinecite{furukawa98}).

If we repeat the considerations for bi-polarons with the respective
optimized size ($L_f \simeq 7$ for the standard parameter set), we find a
critical chemical potential, also depicted in Fig.~\ref{muc_J.eps}, which is
very close to that of polarons.
If we proceed to tri-polarons we find again a similar $\mu^*$.
As can be seen in Fig.~\ref{muc_J.eps}, the chemical potential $\mu^*$
virtually coincides with the chemical potential $\mu^c$ of the
``phase separation'' obtained in Ref.~\onlinecite{KollerPruell2002a}. 
This potential was calculated as the point of coexistence of FM and
AFM order in a homogeneous system.
In view of the rather rough estimate of $\mu^*$ we conclude that
polarons, bi-polarons,  up to phase separated FM regions, are energetically
comparable as long as the size of the FM domains is optimally adapted.
For $J_{\textrm{eff}}\gtrsim 0.12$ which is comparable with the
antiferromagnetic exchange in manganites, individual polarons are
energetically favored.

FM  polarons, however, have a much higher entropy than the other objects and
are therefore thermodynamically favored, even at low temperature.
Moreover, at $T\neq 0$ the domains are not completely (anti)ferromagnetically
aligned which further reduces the energy differences between polaron and
bi-polaron/phase-separated states considerably.
Therefore, even for values of $J_{\textrm{eff}}<0.12$ we find individual
polarons at very low temperatures.
This conclusion is corroborated at $\beta=50$ and $J_{\textrm{eff}}\simeq 0.10$
by the ensuing analysis of MC simulations.

\begin{figure}[h]
  \includegraphics[width=0.33\columnwidth,height=0.52\columnwidth]
  {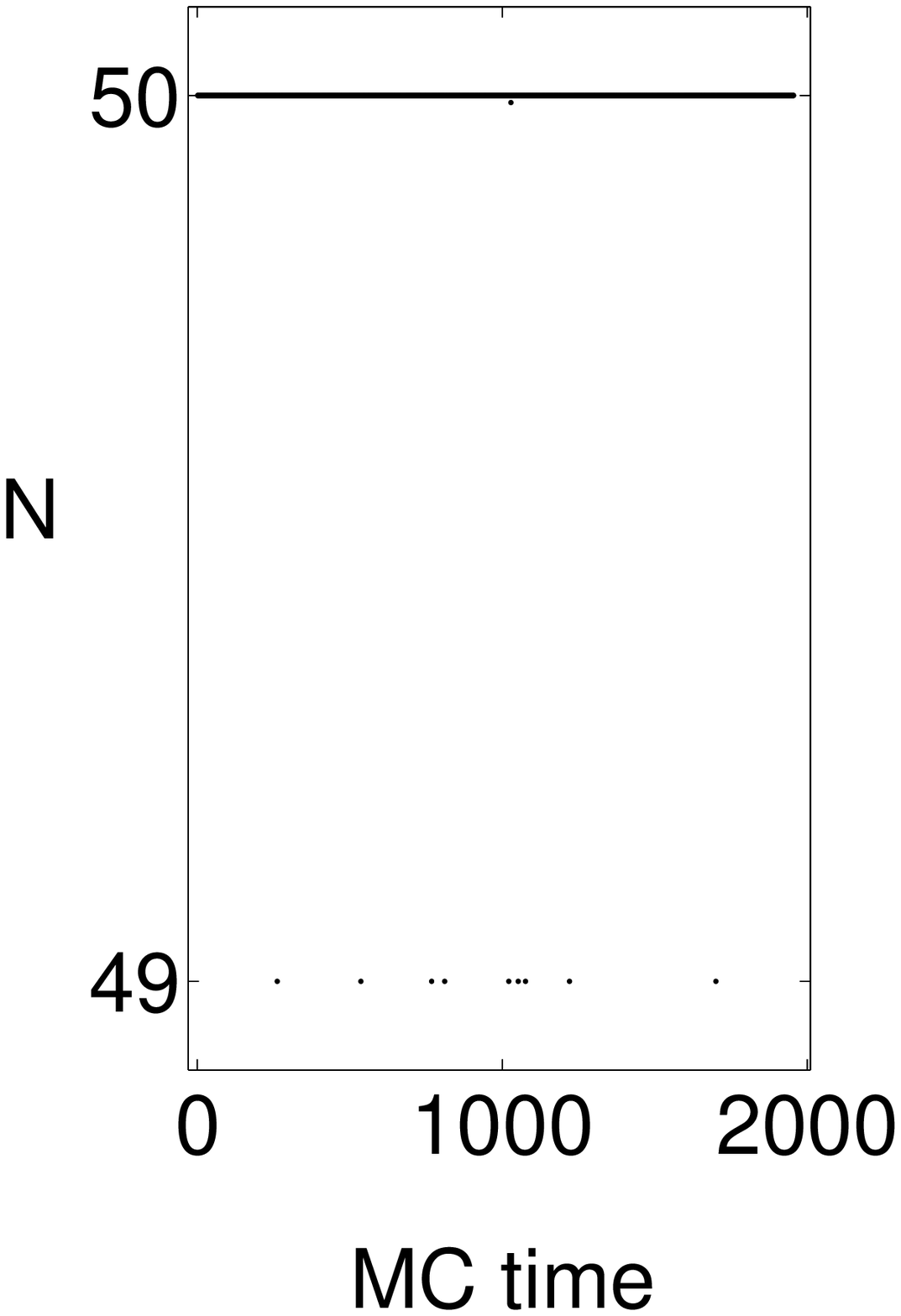}\hfil
  \includegraphics[width=0.33\columnwidth,height=0.52\columnwidth]
  {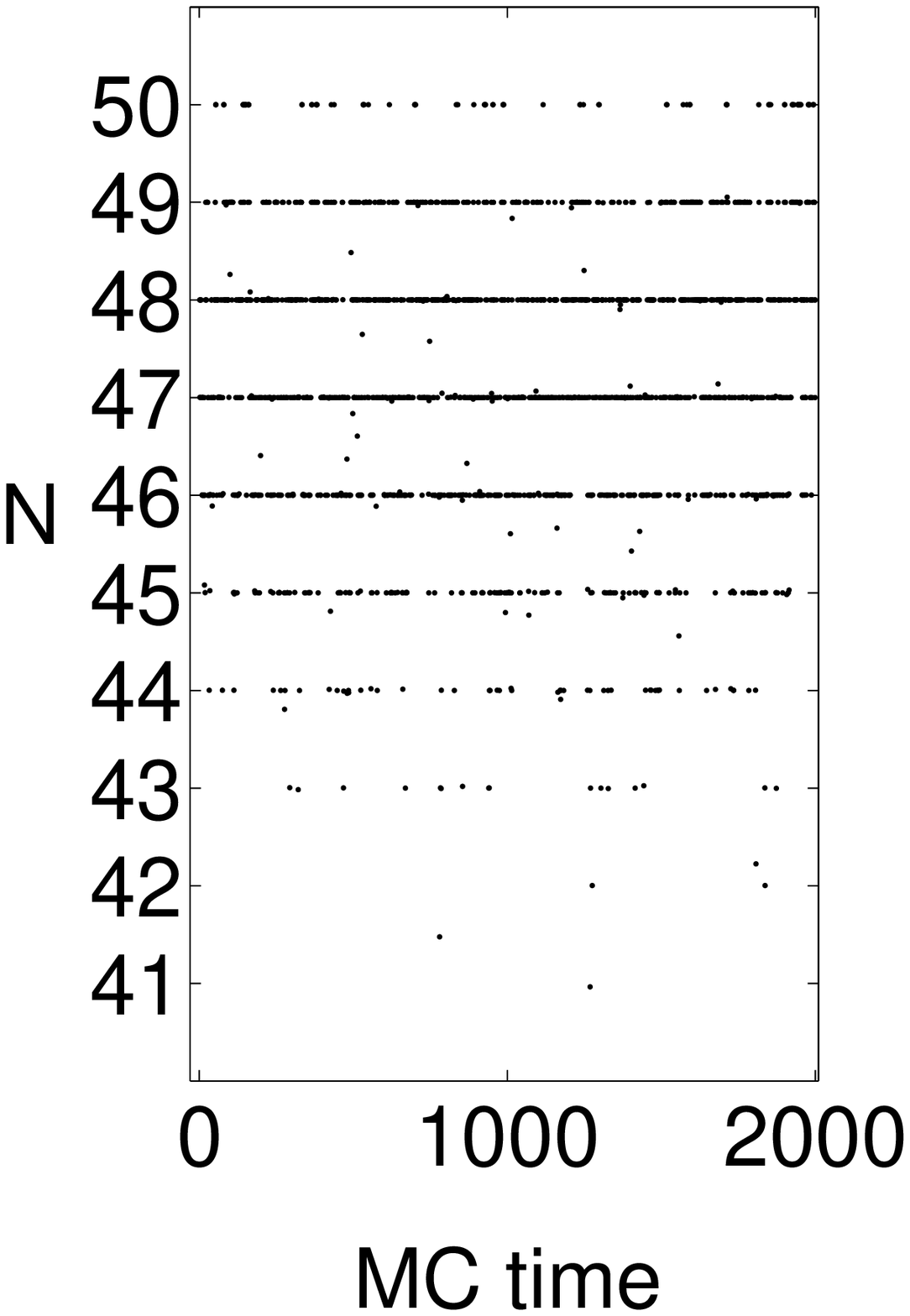}\hfil
  \includegraphics[width=0.33\columnwidth,height=0.52\columnwidth]
  {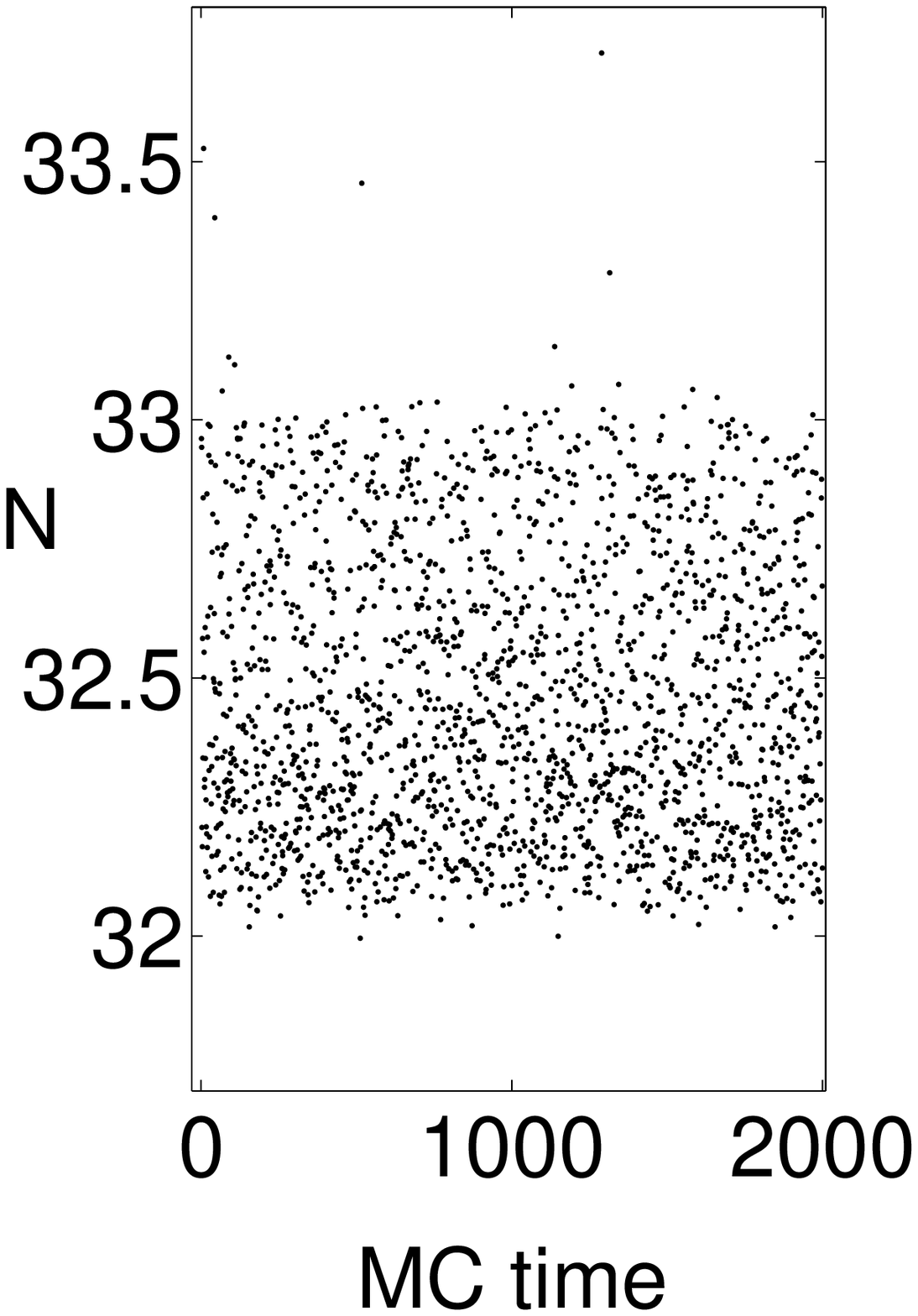}
  \caption{Mean particle numbers in a grand canonical MC simulation ($L=50,
    J_H=6, J'=0.02, \beta=50$) as a function of MC time. One time step
    corresponds to $1000$ sweeps of the lattice.
    a) AFM case ($\mu=1.22$)
    b) polaron regime ($\mu\simeq\mu^*\simeq1.02$)
    c) FM regime ($\mu=0.80$).}
  \label{mean_paricle_numbers}
\end{figure}

In order to scrutinize the polaron arguments, we compute the mean particle
numbers for the corespin configurations entering the Markov chain of
unbiased MC
\[
\langle \hat N \rangle_\SC :=
    \frac{\tr_c\big(\hat N \;\E^{-\beta (\hat H(\SC) -\mu \hat N)}\big)} 
{\tr_c\big(\E^{-\beta (\hat H(\SC) - \mu \hat N)}\big)} =
\sum_{\nu=1}^L \frac{1}{1+\E^{\,\beta(\epsilon_\nu(\SC)-\mu)}}\;,
\]
where $\epsilon_\nu(\SC)$ are the eigenvalues of $\hat H(\SC)$ for the
configuration $\SC$.
As a consequence of the above reasoning
we expect a broad distribution of integer-valued particle numbers
if the chemical potential is close to  $\mu^*$.
The MC time series for $\langle \hat N \rangle_\SC$ for a $L=50$ site chain
($J_H=6, J'=0.02, \beta=50$) is shown in Fig.~\ref{mean_paricle_numbers}.
One time step corresponds to $1000$ sweeps of the lattice.
The inverse temperature $\beta=50$ corresponds to $T\simeq 50-100$~K,
{\it i.e.\ }a temperature relevant for experiments.

The left-hand panel corresponds to a situation where the chemical potential
is far above the critical chemical potential, which has the value
$\mu^*\simeq 1.02$ for the present parameter set.
We see that the band is almost completely filled, with isolated dots at
$N_e=49$ corresponding to occasional FM polarons.
At $\mu^*$ (central panel), in agreement with the polaron picture we find a
broad distribution of {\em integer-valued} mean particle numbers.
If the chemical potential is reduced below $\mu^*$, the system becomes
ferromagnetic and we find the standard result of free electrons with a narrow
and continuous spread in $\langle \hat N \rangle$ not  restricted to integer
values. 

Hole-dressed spin-spin correlations provide another piece of evidence in
favor of FM polarons. 
The bulk of Monte Carlo snapshots (not shown), as taken from simulations for
the FM Kondo model, contains isolated FM polarons of size $L_f=3$ or $L_f=4$.
Once in a while two of them collide and form passing bi-polarons.
The observed fraction of bi-polarons corresponds to a random distribution of
polarons.
\begin{figure}[h]
  \includegraphics[width=0.46\textwidth]{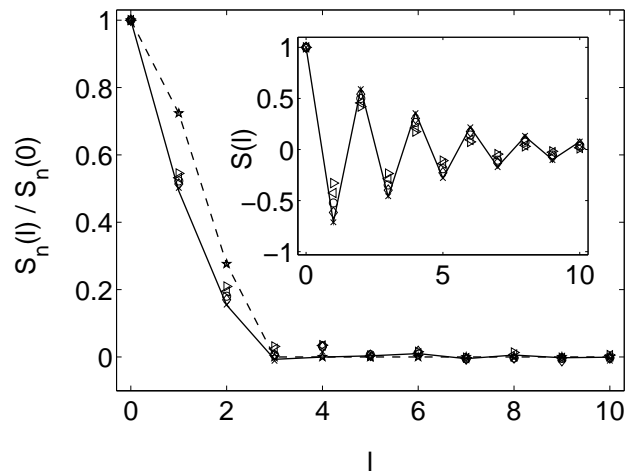}
  \caption{Modified spin-spin correlation function from unbiased MC for an
    $L=50$-site chain containing one ($\times$), two ($\diamond$), three
    ($\circ$), four $(\triangleleft)$, and five $(\triangleright)$ holes.
    The inset shows the conventional spin-spin correlation function
    $  S(l)=\frac{1}{L-l}\,\sum_{i=1}^{L-l} \mathbf{S}_i\cdot
    \mathbf{S}_{i+l}$.
    The dashed line is calculated within the simple polaron picture, while
    the solid line represents the generalized UHA result for a single polaron.
    Parameters are $\beta=50$, $J'=0.02$, and $J_H=6$.}
  \label{css_polarons.eps}
\end{figure}
In order to quantify the information revealed by the MC-snapshots, we introduce
a modified corespin correlation function
\begin{equation}                                              \label{eq:Spol}
  S_n(l)=\frac{1}{L-l}\,\sum_{i=1}^{L-l}
  n_i^h\;\mathbf{S}_i\cdot \mathbf{S}_{i+l}
\end{equation}
that measures the corespin correlations in the vicinity of a charge carrier
(hole). 
The density operator $n_i^h$ for holes at site $i$ is related to the density
operator for electrons via $n_i^h=1-n_i$.
Figure~\ref{css_polarons.eps} shows the results of an unbiased grand
canonical MC simulation.
The observables are evaluated at different subspaces with a fixed particle
number.
We observe ferromagnetic correlations that vanish at $l=3$ corresponding to
a polaron that extends over $L_f=4$ lattice sites.
It should be pointed out that the MC result is almost independent of the
number of holes in the system.
In particular, the data do not indicate any enlargement of the FM domain for
a larger number of holes.
This result can only be explained by individual FM polarons because the size
of the FM domain would strongly increase if there were two or more holes
trapped in it. 

The inset of Fig.~\ref{css_polarons.eps} shows the conventional corespin
correlation function $S(l)$.
We observe the expected antiferromagnetic correlations, which decrease
slightly with increasing number of holes.

The result for the modified spin-spin correlation function can again be
explained qualitatively by the simple polaron picture.
We consider a single polaron in which one charge-carrier is confined.
Let the probability for the hole to be at site $\nu$ in the FM region be
$p_\nu$, which is roughly given by the result for a particle in an infinite
potential well (see Fig.~\ref{polaron.eps}):
$p_\nu \propto \sin^2(\nu\;\pi/(L_f+1))$.
The spin correlation is computed assuming perfect FM order inside
the polaron and perfect AFM order outside.
The result of this simple idea for $L_f=4$ is shown as the dashed line
in Fig.~\ref{css_polarons.eps}.
It agrees qualitatively with the MC data.

For a quantitative, but still fairly simple description, we will now
generalize the uniform hopping approximation, to allow for FM polarons.

\section{UHA for FM polarons}                     \label{sec:UHA_FMP}

In the previous section we have interpreted the MC data by using the
simplest polaron ideas.
In what follows we will refine our polaron picture by including
thermodynamic fluctuations of the corespins.
This is done by a generalization of the finite-temperature
uniform hopping approach (UHA) introduced in
Ref~\onlinecite{KollerPruell2002b}.

In the spirit of UHA, the impact of the corespins on the motion of the
$e_g$ electrons is now described by two UHA parameters,
$u_f$ and $u_a$, for the FM and AFM region, respectively.
These two parameters are averages of the hopping amplitudes in the FM and
AFM domains, respectively.
Their distribution is given by a two-parameter density of states
denoted by $\Gamma_{\!N_f,N_a}(u_f,u_a)$.

The size of individual polarons is fixed to $L_f$ lattice sites.
It is, however, possible that polarons overlap.
The positions of the polarons are specified by the locations
$\{i_1,\ldots,i_m\}$ of their left ends, where $m$ is the number of FM
polaron wells.
If polarons overlap, they may form a bi-polaron
or even greater accumulations of holes.
The grand canonical partition function in this generalized UHA reads
\begin{widetext}
\begin{equation}                                        \label{eq:poL_fart}
  \YY = \sum_{m}\;\sum_{\{i_1,\ldots,i_m\}}\; \iint_0^1 du_f du_a\,
  \Gamma_{\!N_f,N_a}(u_f,u_a)\,
  \tr_c\, \E^{-\beta(\hat  H(u_f,u_a;i_1,\ldots,i_m)-\mu \hat N)}\;.
\end{equation}
\end{widetext}
For 1D chains, subject to open boundary condition, the joint density
$\Gamma_{\!N_f,N_a}(u_f,u_a)$
depends merely upon the number $N_f$ ($N_a$) of bonds in FM  (AFM)  regions.
In higher dimensions, $\Gamma_{\!N_f,N_a}(u_f,u_a)$ would actually depend upon
the location of the individual polarons.
The generalized UHA Hamiltonian reads
\begin{widetext}
\begin{equation}                                        \label{eq:H1d_pol}
  \hat H(u_f,u_a;i_1,\ldots,i_m) = -\sum_{<ij>} \;u_{ij}\;\cdag_i\cnod_j
 - \frac{z}{2 J_H}\,\sum_i \bigg(1-\frac{1}{z}\sum_\delta u_{i,i+\delta}^2\bigg)\;n_i
 + J'\sum_{\langle ij \rangle}(2\;u_{ij}^2-1)\;,
\end{equation}
\end{widetext}
where $\delta$ stands for n.n.\ vectors and $z$ denotes the coordination
number.
The UHA parameter $u_{ij}$ are either $u_f$ or $u_a$, depending upon
the type of magnetic order at the adjacent sites $i$ and $j$.
The integrand of the partition function in \Eq{eq:poL_fart} defines the
joint thermal probability density $p(u_f,u_a\,|\,\beta)$.
From $p(u_f,u_a\,|\,\beta)$, we estimate mean values
$\overline{u}_f,\overline{u}_a$ of $u_f$ and $u_a$ of a $L=20$-site chain
reading
\begin{equation}\label{eq:mean_u}
    \overline{u}_f  = 0.937\;,\quad
    \overline{u}_a  = 0.31
\end{equation}
for the standard parameter set $J'=0.02$ and $J_H=6$ and $\beta=50$.
These mean values are independent of the number of polarons and their
positions as long as the total volume  of the polarons  is small compared to
the system size.
In order to simplify the following discussion, we will replace the thermal
averaging of an observable by the value of that observable at
$\overline{u}_f,\overline{u}_a$.
For the partition function this yields the simple form
\begin{equation}                                        \label{eq:poL_fartII}
  \YY = \sum_{m}\;\sum_{\{i_1,\ldots,i_m\}}\;
  \tr_c\, \E^{-\beta
    (\hat H(\overline{u}_f,\overline{u}_a;i_1,\ldots,i_m)-\mu (L-m))}\;,
\end{equation}
where the influence of the thermal fluctuations of the corespins is contained
in the average hopping amplitudes $\overline{u}_f$ and $\overline{u}_a$.

These average hopping amplitudes, however, are not sufficient
for the determination of observables that do not directly derive from the
partition function such as spin-spin correlations and one-particle spectral
functions.
In principle, these observables can be calculated in UHA by averaging over
a set of {\em typical} thermal corespin configurations $\{\SC\}$ obtained in
UHA. 
In order to construct such a set, 
the azimuthal angles are also required, although they do not enter the energy
and have a flat thermal probability density.
The simplest way of constructing typical corespin configurations is to
draw azimuthal angles at random.
Starting from the reference spin $\mathbf S_\nu$ we proceed to the
neighboring corespins by adding a random azimuthal angle $\chi$ to the fixed
relative polar angle. Thus we obtain a collection of typical corespin
configurations $\{\SC\}$.

\subsection{Static Correlations}

We continue the discussion of the modified spin-spin correlation function
$S_n(l)$.
For a quantitative, but still fairly simple description, we take the deviations
from perfect FM and AFM order into account, while for the hole the
approximate probabilities $p_\nu$ are retained.
We employ the mean UHA parameters $\overline{u}_f$ and $\overline{u}_a$ to
describe the relative angles of neighboring corespins and average over
typical corespin configurations with random hole positions.
A comparison of the unbiased MC results with those of this approximate
polaron approach, depicted in Fig.~\ref{css_polarons.eps}, reveals an
excellent agreement.

\begin{figure}[h]
  \includegraphics[width=0.90\columnwidth]{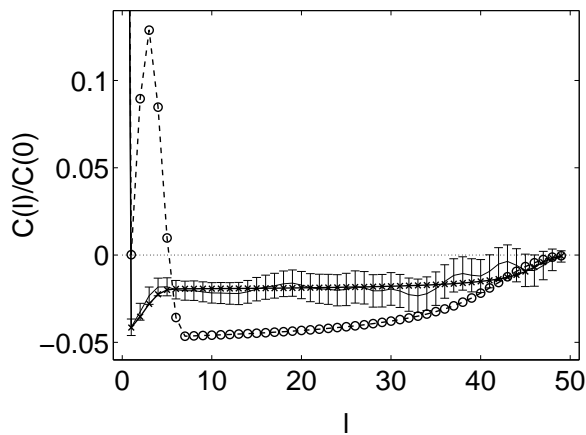}
  \caption{Density density correlation function for  $L=50, \beta=50, J'=0.01,
    J_H=6$, and $N_h=2$.
    Error bars represent unbiased MC data, crosses stand for polaron-, and
    circles for bi-polaron results in UHA.  }
  \label{fig:CF}
\end{figure}

Another observable to distinguish between polarons, bi-polarons, or even
phase separated high density FM clusters is the density-density correlation
function 
\begin{equation}\label{eq:nSS}
C(l) := \frac{1}{L-l}\sum_{i=1}^{L-l} \;\langle (n^h_i-\langle n^h_i\rangle)
(n^h_{i+l}-\langle n^h_{i+l}\rangle)  \rangle\;.
\end{equation}
If holes form independent FM polarons, the correlation function should be
structureless, while if holes gather in one FM regime, the correlation
function will exhibit a positive peak at a typical inter-particle distance.
In Fig.~\ref{fig:CF} $C(l)$ is shown for the two-hole subspace, where
only those spin configurations of the Markov chain are taken into account,
for which $N_h\simeq 2$.
The UHA-polaron result is derived as follows.
The positions of two FM-regions of size $L_f=4$ are chosen at random,
including overlapping ones.
The hopping parameters are $\bar u_f$ ($\bar u_a$) for FM (AFM) bonds and the
resulting tight-binding model is solved.
The lowest two eigenstates will then be localized in the two FM potential
wells. 
The resulting correlation functions are averaged over all possible positions
of the FM potential wells.
We observe a strikingly close agreement with the unbiased MC results.
Similarly we proceed in the bi-polaron case, which is characterized by a
single FM region of optimized size ($L_f=7$).
Here the two holes occupy the ground state and first excited
state of the FM potential well.
The resulting correlation function differs drastically from the MC
data.

This discrepancy increases with increasing hole number, which shows clearly
that the physics of the 1D FM Kondo model is correctly described by
single-hole polarons and not by phase separation.

\subsection{Polaronic features in the spectral density}\label{sec:sf}

In this subsection, we compute the spin-integrated spectral density
\[
A_k(\omega)=-\frac{1}{\pi} \sum_\sigma\;\Im
 \ll a\phd_{k \sigma}; a^\dagger_{k \sigma} \gg_\omega
\]
for the original DE model by unbiased MC simulations based on the expression
\Eq{eq:MC_GF2}.
The results will again be analyzed by UHA in the framework of the polaron
ansatz.
In particular the pseudogap in the spectral density near $n=1$, found in the
FM Kondo model~\cite{moreo99,KollerPruell2002a}, can readily be
explained in the polaron picture.
It is a consequence of the ferromagnetic box in which the hole moves.

The computation of $A_k(\omega)$  in UHA is based on the reasoning
that led to the partition function in \Eq{eq:poL_fartII}.
I.e., for each polaron configuration and each value of $\bar u_a$, eigenvalues
$\eps^{(\lambda)}$ and eigenvectors $|\psi^{(\lambda)}\rangle$ of the respective
tight-binding Hamiltonian are determined, from which the Green's function
\[
\ll c\phd_{i}; c^\dagger_{j} \gg^{\bar u_a,\bar u_f}_\omega\;.
\]
is determined in local spin-quantization.
The transformation to the global spin-quantization is given by
\Eq{eq:MC_GF2}
\begin{equation}\label{eq:trafo_local_global}
\sum_\sigma \ll a\phd_{i \sigma}; a^\dagger_{j \sigma} \gg^\SC_\omega
    = \;u^{\UA\UA}_{j,i}(\SC)
    \ll c\phd_{i}; c^\dagger_{j} \gg^{\bar u_a,\bar u_f}_\omega\;.
\end{equation}
In the framework of UHA the relative angles of neighboring spins
are fixed by the parameters $\bar u_a$ and $\bar u_f$.
For a unique description of the entire spin configuration, however, azimuthal
angles are again required.
We proceed like in the discussion corresponding to \Eq{eq:nSS}, i.e.\ for
fixed parameters ($\bar u_f, \bar u_a$), spin configurations are generated
with a flat sampling distribution in azimuthal angles.
This is little numerical effort, as only the prefactor $u^{\UA\UA}_{ij}(\SC)$
is affected.

In the transition region ($\mu\simeq \mu^*$), the number of polaron wells is
not well defined, as pointed out in conjunction with
Fig.~\ref{mean_paricle_numbers}.
In order to obtain a detailed understanding we compare unbiased MC data and
results of the polaron ansatz in the subspace of fixed particle (polaron)
number.

\subsubsection{Antiferromagnetism at half filling}

We begin the discussion with the spectrum of the AFM state for
the completely filled lower Kondo band (no polarons).
\begin{figure}[h]
  \psfrag{pi }{$\pi$}
  \psfrag{ 0 }{$0\,$}
  \psfrag{omega}{$\omega$}
  \includegraphics[width=0.48\columnwidth,height=0.5\columnwidth]
  {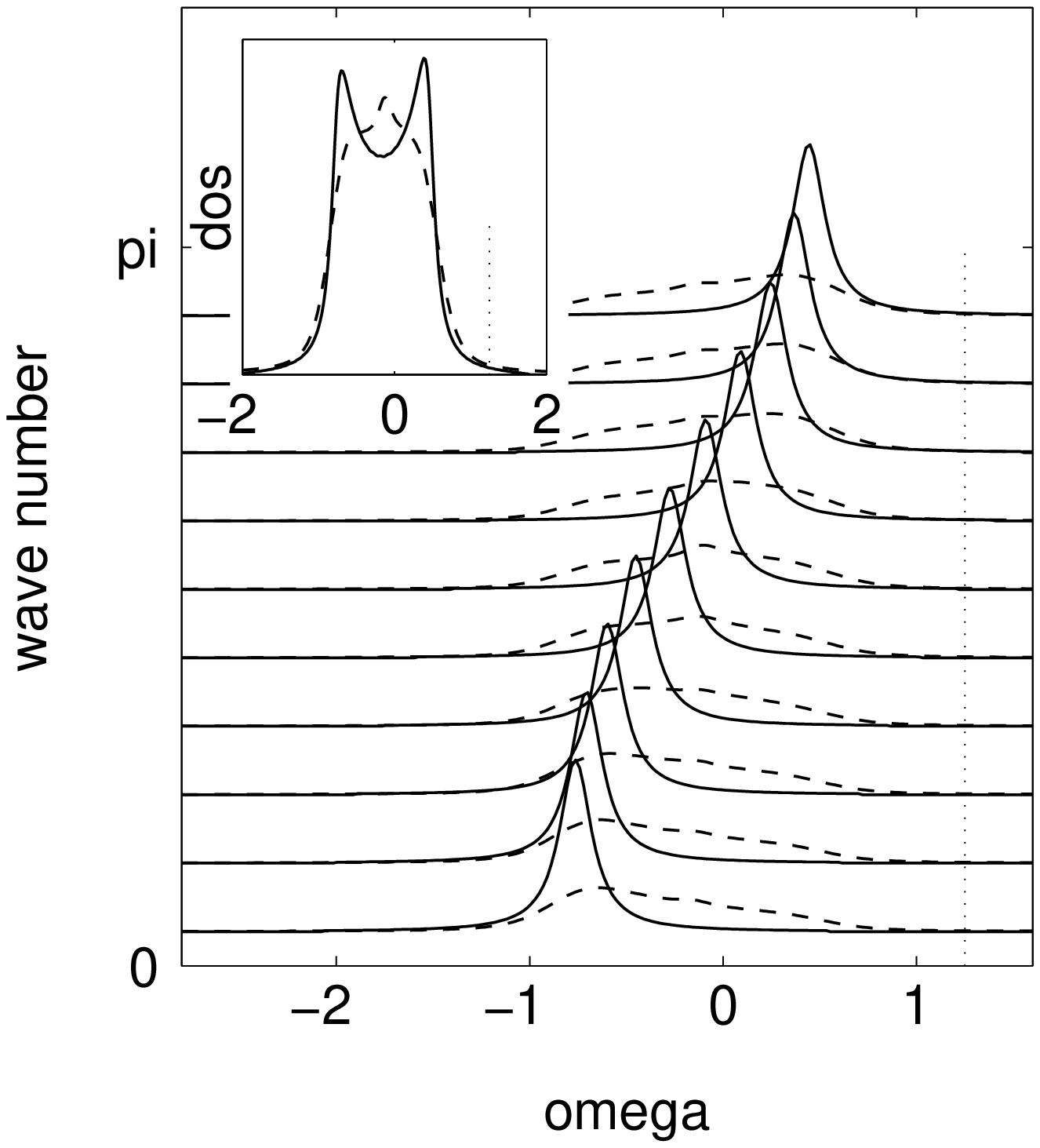}\hfil
  \includegraphics[width=0.48\columnwidth,height=0.5\columnwidth]
  {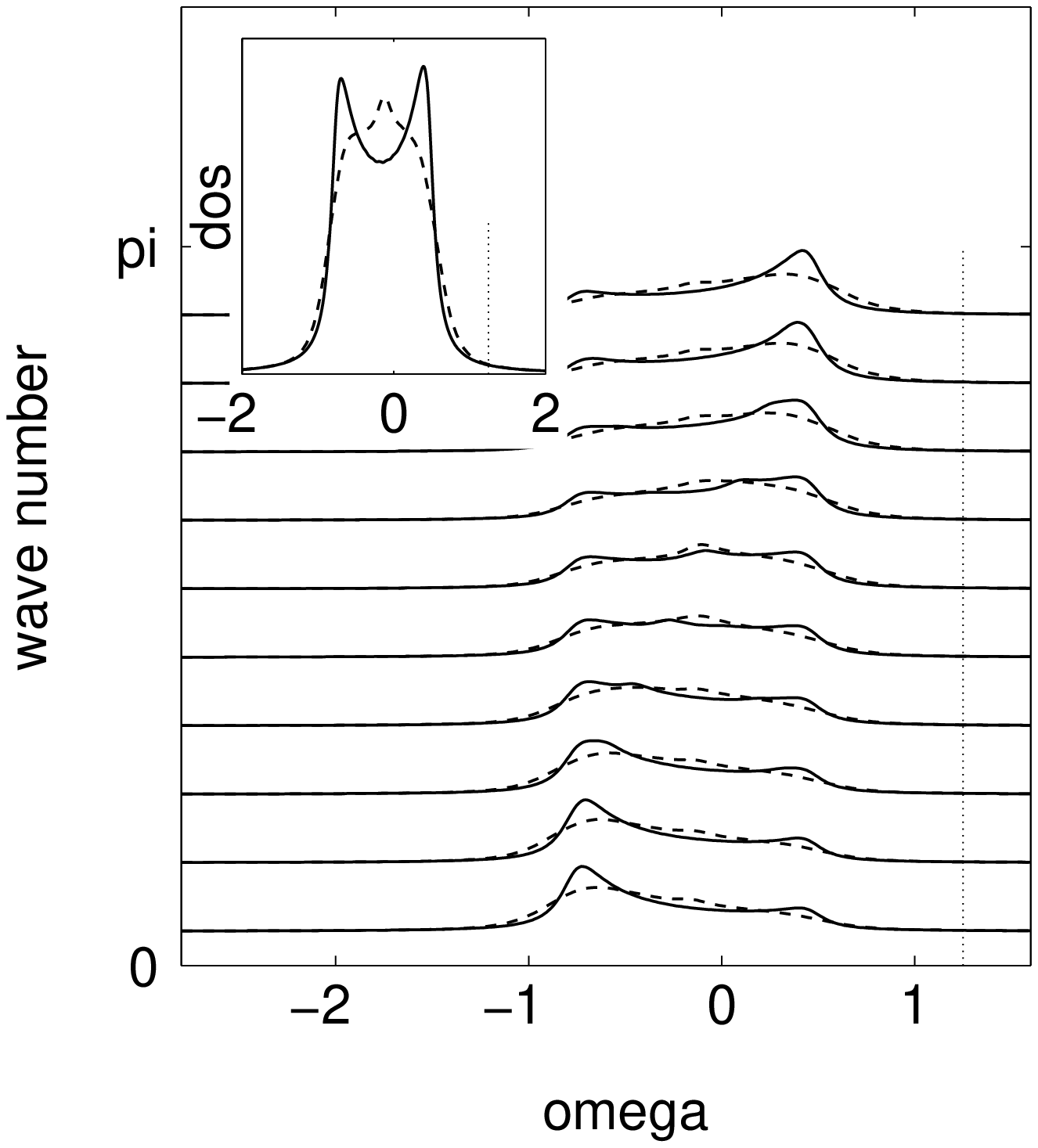}
  \caption{Spectral density of a half-filled Kondo chain
    with parameters $L=20, \beta=50, J'=0.02, J_H=6$, and $\mu=1.25$.
    Dashed lines represent unbiased MC data, solid lines stand for
    UHA result ($\bar u_a = 0.31$). 
    Inset: density of states.
    The vertical bar indicates the chemical potential. 
    The left(right)-hand panel shows UHA-results in local (global) spin
    quantization. 
    Error bars have been omitted for clarity.
   }
  \label{sw_L20N20_beta50.eps}
\end{figure}
The unbiased MC data as shown in Fig.~\ref{sw_L20N20_beta50.eps}
display broad structures due to incoherent motion of charge
carriers in a spin background that exhibits random deviations from perfect
AFM order.
The result can be described by the UHA ansatz.
For the mean value of the hopping amplitude we have $\bar u_a=0.31$
which corresponds to
$\langle S_i S_{i+1}\rangle = \cos(\vartheta) \simeq -0.7$.
Since there are no holes, and consequently no polarons, the UHA parameters
are the same for all bonds.
The left-hand panel in Fig.~\ref{sw_L20N20_beta50.eps} shows the UHA result
in local quantization, i.e.\ without the transformation given in
\Eq{eq:trafo_local_global}.
In local quantization, the spectral density is simply given by
\[
A_k(\omega) = \delta(\omega - 2 \bar u_a \cos(k))\;.
\]
The agreement with the unbiased MC result is rather poor at this stage
although the band width is already well approximated.
In the MC spectra, only very weak remnants of the tight-binding features are
visible on top of the incoherent spectrum, which is almost k-independent.

However, if we take into account the necessary transformation
\Eq{eq:trafo_local_global} to a global spin quantization,
the agreement is strikingly close (see right-hand panel).
The quasi-particle dispersion is strongly smeared out due to
the random azimuthal angles of the corespins.
There are two minor discrepancies between UHA and unbiased MC.
The tight-binding remnants are more pronounced in UHA, while the MC results
exhibit a weak structure near $\omega=0$, which is due to random fluctuations
in the relative n.n.\ angles, resulting in locally trapped electrons.
Nonetheless, at half-filling UHA and unbiased MC simulations yield compatible
results for the spectral density and the density of states
(insets of Fig.~\ref{sw_L20N20_beta50.eps}).

\subsubsection{Ferromagnetic Polarons}
\begin{figure}[h]
  \psfrag{pi }{$\pi$}
  \psfrag{omega}{$\omega$}
  \includegraphics[width=0.9\columnwidth]{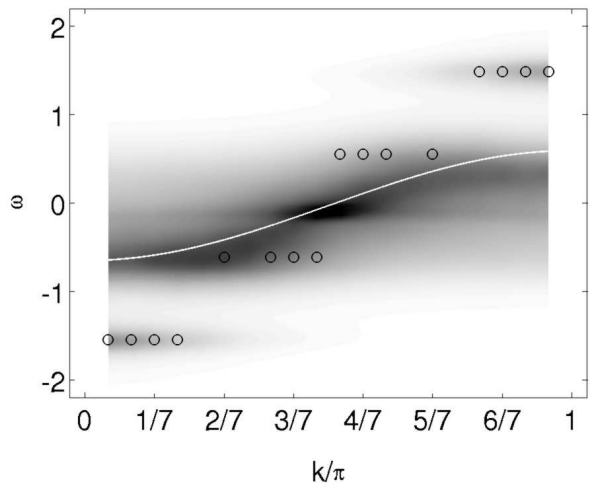}
  \caption{Spectral density for $N_h=1$ holes (one polaron).
    Parameters like in Fig.~\ref{sw_L20N20_beta50.eps}, except $\mu=1.1$.
    Comparison of MC data (gray scale plot) with results of the
    simple polaron-well model.
    Circles (solid bright lines) stem from states localized in the FM
    (AFM)-well.
    Hopping parameters for the polaron-well model are $\bar u_f=0.937$,
    $\bar u_a=0.31$, and $L_f=4$.}
  \label{Ak_grayscale_P1.eps}
\end{figure}

\begin{figure}[h]
  \psfrag{pi }{\Large $\pi$}
  \psfrag{omega}{\Large $\quad \omega$}
  \includegraphics[width=0.8\columnwidth]{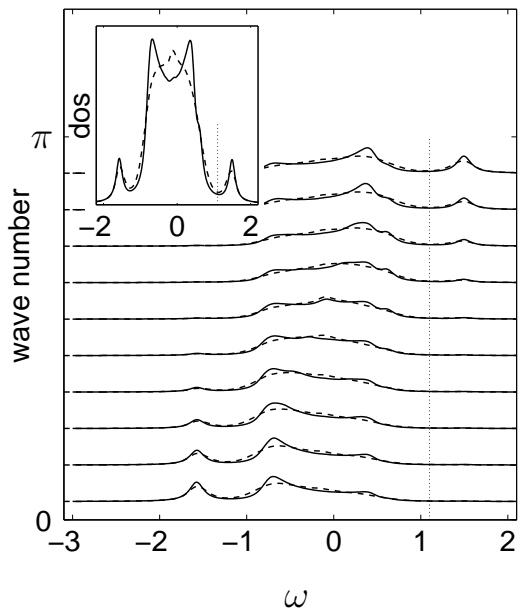} 
  \caption{Spectral density for $N_h=1$ hole (one polaron).
    Symbols like in Fig.~\ref{sw_L20N20_beta50.eps} and parameters like in
    Fig.~\ref{Ak_grayscale_P1.eps}.
    The broad incoherent central part of the spectrum
    (width $\simeq 2\bar u_a$) derives from the
    motion of the electrons in the fluctuating AFM background.
    Polaronic peaks show up at $\omega \simeq \pm 1.5$.
    In between these two structures a pseudogap opens at the chemical
    potential $\mu^*$. It is accompanied by a mirror pseudogap near $-\mu^*$.}
  \label{sw_L20N19_beta50.eps}
\end{figure}

Next we consider the case of one hole in the otherwise half-filled Kondo chain.
To this end, we investigate the grand canonical MC data in the $N_h=1$ subspace.
The respective MC spectrum is shown in Figs.~\ref{Ak_grayscale_P1.eps}
and~\ref{sw_L20N19_beta50.eps}.
The main feature of the spectrum is a broad incoherent background, similar
to the one found in the AFM case.
In addition, two dispersionless structures are visible at $\omega\approx\pm
1.5$.
As discussed earlier~\cite{moreo99,KollerPruell2002a}, a pseudogap shows up
at the chemical potential.
We find that an additional (mirror) gap appears at the opposite side of the
spectrum.

Before discussing the UHA result, we want to provide a rough explanation
of the MC data in terms of a simple polaron model.
We assume an FM polaron-well of size $L_f=4$, characterized by a tight-binding
hopping parameter $\bar u_f$ embedded in an antiferromagnetic background with
hopping parameter $\bar u_a$.
Since the two hopping parameters are very different,
we treat the various regions as separate chains and neglect their interaction.
I.e.\ there is an isolated tight-binding chain of size $L_f$ corresponding to
the FM region, and one or two chains corresponding to the AFM background.
A carrier is localized either in the FM or the AFM domain.
The eigenvalues in the FM region, $\varepsilon^f_\nu=-2 \bar u_f\cos(k^f_\nu)$,
depend on the momentum $k^f_\nu=\nu\pi/(L_f+1)$ with $\nu=1,\ldots,L_f$.
The energies of states corresponding to different localizations of the FM
well ($i_0=1,\ldots,L-L_f$) are degenerate.
The corresponding eigenstates in real space read
\[
\psi^{(k_\nu,i_0)}_i\propto
\left\{%
\begin{array}{ll}
    \sin((i-i_0+1) k_\nu), & \hbox{$i_0\le i < i_0+L_f$} \\
    0, & \hbox{otherwise.} \\
\end{array}%
\right.
\]
In this simple polaron-well model the expression for that part of the
spectral density in local spin-quantization is
\begin{equation}                                    \label{eq:Akw2}
  \begin{aligned}
  A^f_k(\omega) &\propto
 \sum_{\nu=1}^{L_f}\;\;
  \sum_{i_0=1}^{L-L_f}
  \big| \langle k|\psi^{(k_\nu,i_0)}\rangle\big|^2
  \;\delta(\omega-\eps^f_\nu)\\
  &\propto \sum_{\nu=1}^{L_f}\;c_\nu(k)\; \delta(\omega-\eps^f_\nu)\;,
  \end{aligned}
\end{equation}
where $| k\rangle$ stands for the eigenvectors of the homogeneous tight-binding
model with open boundary condition, i.e.\ $\langle j |k\rangle \propto
\sin(j\, k)$.
The coefficient $c_\nu(k)$ as function of $k$ shows a broad hump at $k\simeq
k_\nu$. Hence, the contributions of the FM regions to $A_k(\omega)$ are
dispersionless structures at energies $\omega=\eps^f_\nu$ which are
concentrated about $k=k^f_\nu$.
These structures are marked by open circles in the grayscale plot of
Fig.~\ref{Ak_grayscale_P1.eps}.
They explain the additional features at the band edges, which are clearly
visible in the spectral density in addition to the broad incoherent background.
The latter is due to the motion of the hole in the AFM regions.
Since the AFM regions are much larger than the FM well, a continuous
tight-binding band develops, characterized by the hopping parameter $\bar
u_a$. The band is shown as a white line in Fig.~\ref{Ak_grayscale_P1.eps}.
This part of the spectrum is similar to that at half filling
(Fig.~\ref{sw_L20N20_beta50.eps}).
The remaining discrepancy as compared to the MC data is due to the
fluctuations of the azimuthal angles of the corespins causing the white
line to become more incoherent, as shown by the UHA calculations below.
The transformation from local to global spin-quantization has, however,
negligible impact on the polaron states in the spectrum, since they are due
to the FM region, in which the fluctuations of the corespins are less relevant.
As we can see in Fig.~\ref{Ak_grayscale_P1.eps}, our reasoning based on a
single polaron well already describes the qualitative features correctly.

The origin of the pseudogap and its 'mirror image' on the opposite side of
the spectrum can now be simply identified as the energy difference between the
uppermost (lowest) state in the FM potential well $(E = \mp 2 \bar u_f
\cos(\pi/(L_f+1)))$ and the upper (lower) edge of the tight-binding band in the
AFM region $(E = \mp 2 \bar u_a)$ leading to a width of the pseudogap $\Delta
E = 2(\bar u_f \cos(\pi/(L_f+1))-\bar u_a)$, see Fig.~\ref{Ak_grayscale_P1.eps}.

A different picture would emerge if the FM domains were more extended, as
for example in a PS scenario.
They would then contain many energy levels
(not only four as in Fig.~\ref{Ak_grayscale_P1.eps}) and should thus
{\it  not} give rise to a pseudogap.

For a more quantitative description we invoke the two-parameter UHA as
described before. 
In Fig.~\ref{sw_L20N19_beta50.eps} the UHA results, already in global
quantization, are compared with those of unbiased MC simulations.
The features of the spectral density are well reproduced.
UHA even yields quantitative agreement as far as the pseudogap in the
density of state is concerned.

\begin{figure}[h]
  \psfrag{pi }{$\pi$}
  \psfrag{omega}{$\omega$}
  \includegraphics[width=0.48\columnwidth,height=0.5\columnwidth]
  {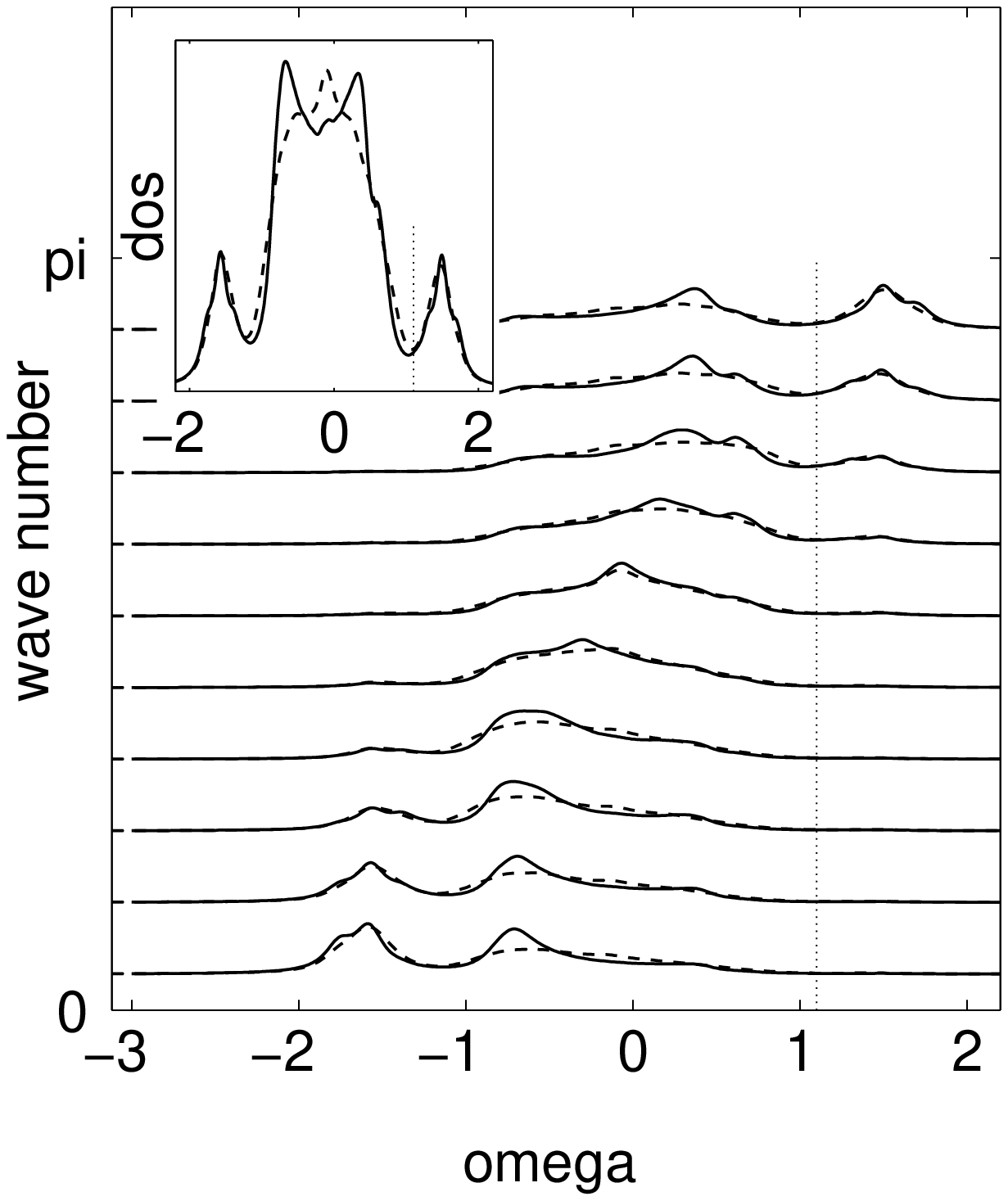}
  \hfil
  \includegraphics[width=0.48\columnwidth,height=0.5\columnwidth]
  {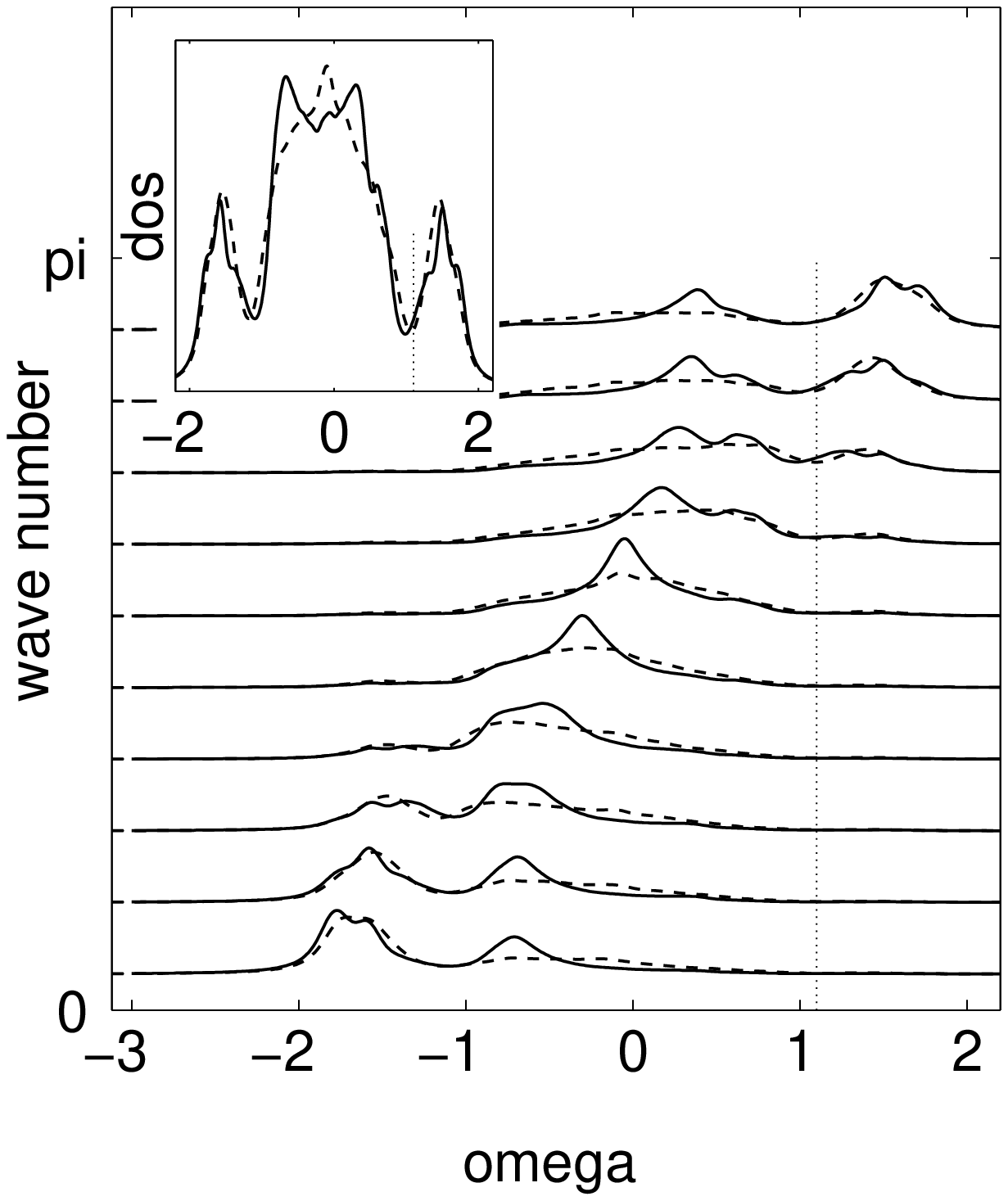}
  \caption{Spectral density for $N_h=2$ holes (two polarons) in the left-hand
  panel and $N_h=3$ holes (three polarons) in the right-hand panel.
    Parameters and meaning of symbols like in Fig.~\ref{sw_L20N19_beta50.eps}.}
  \label{sw_L20N18_beta50.eps}
\end{figure}

For the same value of the chemical potential $\mu=1.1$, the two- and three-
hole subspaces have been studied by projecting the grand canonical MC data on the $N_h=2$ and
$N_h=3$ subspace, respectively.
The results are depicted in Fig.~\ref{sw_L20N18_beta50.eps}.
The spectra are qualitatively similar to those for the one-hole subspace.
Only the spectral weight of the polaron peak increases and
shoulders show up in the AFM part of the spectrum.
They are due to the interstitial AFM regions enclosed by the FM-polaron
wells.
The argument is the same as before.
The allowed energies are $\varepsilon(k) = -2 \bar u_a \cos(k)$, but now the
possible $k$-values depend on the size of the interstitial regions.

\subsection{Discussion}

The emerging global picture is as follows.
There exists a critical chemical potential~$\mu^*$.
The value of $\mu^*$ can be obtained from simplified energy considerations.
For $\mu$ significantly above $\mu^*$, the band is completely filled
and antiferromagnetic.
The spectral density in this case is shown in Fig.~\ref{sw_L20N20_beta50.eps}.
At $\mu^*$, according to Fig.~\ref{mean_paricle_numbers},
holes enter the $e_g$-band forming isolated FM domains each containing a
{\em  single} hole (see Figs.~\ref{css_polarons.eps} and \ref{fig:CF}).
In the grand canonical ensemble, the number of polarons strongly fluctuates
and the height of the polaron peak in the spectrum is directly linked to the
number of holes.
The pseudogap appears around the critical chemical potential.
For values below $\mu^*$, the system switches from predominantly AFM order to
FM behavior and the pseudogap gradually disappears in favor of a
single quasi particle band of tight-binding type.
Our analysis yields compelling evidence against the PS scenario and in favor
of FM polarons in 1D.
Furthermore, it appears plausible that the formation of FM polarons will
exist in symbiosis with lattice deformations (Jahn-Teller
polarons)~\cite{EdwardsI,millis96:polaronsI}. 

For CMR oxides, Saitoh {\it et al\ }~\cite{DessauII} have investigated
the temperature dependence of angle resolved photoemission spectra
(ARPES) for the phase transition from FM to PM order. 
These studies show that a pseudogap also develops above $T_C$,
which can be rationalized in the polaron picture.
In the PM phase we have the competition of ferromagnetism,
driven by the DE mechanism, and spin disorder due to thermal fluctuations.
Therefore, FM polarons will form in the paramagnetic background.
They will, however, be more extended because the PM force is less pronounced
than the AFM force at low temperatures\cite{horsch99}.
With increasing temperature, the corespin fluctuations become stronger
and the competition of the FM polarons with the PM background gets tougher.
The existence of FM domains above $T_C$ has been 
corroborated by neutron scattering experiments~\cite{Perring97}.

Furthermore, the ARPES experiments revealed that the bandwidth changes merely
by about $4\%$ across the FM to PM phase transition.
On the other hand, it has been argued~\cite{DessauII} that the DE model
predicts a reduction of about $30\%$ in the PM phase if there the mean angle
between neighboring spins is taken to be $\pi/4$ and the mean hopping
parameter is therefore reduced to $1/\sqrt{2}$. 
The authors in Ref.~\onlinecite{DessauII} therefore conclude that
'DE is probably not even the dominant mechanism \ldots'.

At first glance, the argument seems convincing.
However, in the polaron picture we do not really expect such a
dramatic change of the band width since it is determined by the
polaronic peaks at $E=\pm 2 \bar u_f\cos(\pi/(L_f+1))$.
Hence, the band edges depend on the hopping parameter of the FM region and
not on that of the PM region.
Moreover, by the same reasoning that leads to a bandwidth reduction of
$30\%$ in the PM state, one would expect that the bandwidth vanishes in the
AFM phase at low temperatures, since here the neighboring corespins are
mostly antiparallel.
That conclusion is in strong contrast to the unbiased MC results depicted
in Figs.~\ref{sw_L20N19_beta50.eps} and~\ref{sw_L20N18_beta50.eps}. 
Even for the incoherent inner part of the spectrum, which is due to electronic
motion in the AFM region, a considerable band width exists, due to the spin
fluctuations which are present even at very low temperatures. 
On top of that, the band-edges at finite hole filling are not really
determined by the AFM regime, but rather by the FM polarons. 

\section{Conclusions}                           \label{sec:conclusion}

In this paper, polaronic aspects of the ferromagnetic Kondo
(double-exchange) model have been analyzed by unbiased finite temperature
Monte-Carlo simulations and they have been explained by simple physical pictures.
It has been found that in 1D, the physical effects of the FM Kondo model
close to half filling are not governed by phase separation, as previously
reported, but rather by single-hole ferromagnetic polarons.
They can be explained qualitatively on the back of an envelope by idealized
polaron pictures.

It seems sensible to reassess the explanations of CMR based on PS.
These explanations are primarily based on percolation ideas, which can
equally well be applied to FM polarons as percolating units.
It appears plausible that the formation of FM polarons will exist in
symbiosis with lattice deformations (Jahn-Teller polarons).
Single-hole FM polarons allow a direct explanation of the pseudogap,
observed in the manganites, whereas for larger FM clusters the pseudogap
would be filled up by additional states.
The striking similarity of the bandwidth of the FM and the PM phase,
observed in ARPES experiments, can also be explained by FM polarons in the
frame of the DE model.
Moreover, the infinite compressibility near the half filled band, which has
previously been attributed to PS, is a consequence of the fluctuating number
of polarons in the grand canonical ensemble.
Work is in progress for higher-dimensional systems where the entropy is
expected to have less influence on the thermodynamic behavior.

For the analysis of the Monte Carlo results,
we have extended the uniform hopping approach (UHA) at finite
temperatures to include polaronic effects.
This ansatz reduces the numerical effort by several orders
of magnitude, while retaining all crucial physical features.
The key idea is to map the physics of the high-dimensional configuration
space of the $t_{2g}$ corespins onto an effective two-parametric model.
A full thermodynamic evaluation of the UHA model takes into account entropy
and fluctuations of the corespins. 
The results are in close agreement with the unbiased MC data
and allow a realistic description of all FM polaron effects found in various
physical quantities.

\section{Acknowledgement}

This work has been supported by the Austrian Science Fund (FWF), project
no.\ P15834-PHY. We thank E. Dagotto for useful comments.

%
%

\end{document}